\documentclass[cits]{PoS}

\usepackage{slashed,amsmath}

\usepackage{dsfont}

\title{Scale-dependence of the $B$-meson LCDA beyond leading order from conformal symmetry}

\ShortTitle{ $B$-meson LCDA from conformal symmetry}

\author{{Vladimir M. Braun}
\\
        Institut f\"ur Theoretische Physik, Universit\"at
   Regensburg, D-93040 Regensburg, Germany\\
        E-mail: \email{vladimir.braun@ur.de}}
\author{{Yao Ji}
\\
        Institut f\"ur Theoretische Physik, Universit\"at
   Regensburg, D-93040 Regensburg, Germany
\\
Theoretische Physik 1, Naturwissenschaftlich-Technische Fakult{\"a}t,
Universit{\"a}t Siegen, D-57068 Siegen, Germany
\\
        E-mail: \email{yao.ji@ur.de}}
\author{\speaker{Alexander N. Manashov}\\
II. Institut f{\"u}r Theoretische Physik, Universit{\"a}t Hamburg, D-22761 Hamburg, Germany\\
       St.Petersburg Department of Steklov Mathematical Institute, 191023 Saint-Petersburg, Russia \\
       E-mail: \email{alexander.manashov@desy.de}}

\abstract{
We argue that the evolution kernel for the scale-dependence of the $B$-meson light-cone distribution
amplitude (LCDA) can be written, to all orders in perturbation theory, in terms of the generator
of special conformal transformations in a modified theory: QCD at critical coupling in non-integer $d\slashed{=}4$
dimensions.
Explicit expression for the eigenfunctions of the evolution kernel is derived that is valid, again, to
all orders. From a practitioner's point of view the utility of this representation is that it allows
to ``save one loop'' and obtain the evolution kernel to a given order of perturbation theory,
up to a constant term, from the calculation of the
conformal anomaly at one order less.
This construction is verified by explicit calculation at two-loop level.
}

\FullConference{14th International Symposium on Radiative Corrections (RADCOR2019)\\
9-13 September 2019\\
		Palais des Papes, Avignon, France }

\begin{document}

\section{Introduction}
The $B$-meson light-cone distribution amplitude (LCDA) plays a central role in QCD description of
$B$-decays with final states involving energetic light hadrons. Such processes deliver important
information on the structure of the flavor sector in the Standard Model and may also open a window
to new physics. It is therefore not surprising that their theoretical and experimental study
is attracting continuous interest, see e.g.~\cite{Kou:2018nap}.

The leading-twist LCDA considered in this talk was first introduced in~\cite{Grozin:1996pq}
as a matrix element between the vacuum and $B$-meson state of the renormalized light-ray operator
\begin{align}\label{Oz}
\mathcal{O}(z)&=[\bar q(zn) \slashed{n}\gamma_5 h_v(0)]\,,
\notag\\
\langle 0| \mathcal{O}(z)|\bar B(v)\rangle &=
i F(\mu) \Phi_+(z,\mu)
=
i F(\mu) \int_0^\infty\!d\omega\, e^{-i\omega z} \phi_+(\omega,\mu)\,.
\end{align}
Here $h_v(0)$ is an HQET effective heavy quark  field  and $\bar q(zn)$ is a light anti-quark,
$v_\mu$ is the heavy quark velocity, and $n_\mu$ is a light-like vector, $n^2=0$, such  that $n\cdot v=1$.
The Wilson line connecting the fields is tacitly implied.
The square brackets in the definition of the light-ray operator stand for renormalization in the
$\overline{\text{MS}}$ scheme, $\mu = \mu_{\scriptscriptstyle \overline {\rm MS}}$ is the factorization scale
and $F(\mu)$ is the HQET $B$-meson decay constant~\cite{Neubert:1993mb}.
One can show that the LCDA in position space, $\Phi_+(z,\mu)$,
is an analytic function of $z$ in the lower complex half-plane.
The scale-dependence of the LCDA is inherited from the renormalization-group (RG) equation
for the nonlocal operator $\mathcal O(z)$  which  has a standard form
\begin{align}\label{RGE}
 \Big(\mu{\partial_\mu}+\beta(a){\partial_a}+\mathcal{H}(a)\Big)
\mathcal O(z)=0\,.
\end{align}
Here $\partial_\mu = \partial/\partial\mu$, etc., $\beta(a)$ is the QCD beta-function,
$a=\alpha_s/4\pi$ is the strong coupling and $\mathcal{H}(a)=a\mathcal{H}^{(1)}+a^2 \mathcal H^{(2)}+\ldots$
is a certain integral operator (evolution kernel).

The one-loop kernel $\mathcal H^{(1)}$ was first calculated by Lange and
Neubert~\cite{Lange:2003ff}.
The result in position space reads~\cite{Braun:2003wx,Knodlseder:2011gc}
\begin{align}\label{Honeloop}
\mathcal H^{(1)} \mathcal O(z) &=4C_F\biggl\{ \left(\ln(i \mu_{\scriptscriptstyle\overline{\text{MS}}}\,
e^{\gamma_E} z)-\frac14\right)\mathcal O(z)
+\int_0^1 d\alpha\frac{\bar\alpha}\alpha\Big(\mathcal{O}(z)-\mathcal O(\bar\alpha z)\Big)\biggr\},
\end{align}
where $\bar\alpha = 1-\alpha$.
It is possible to show \cite{Knodlseder:2011gc, Braun:2014owa}
that $\mathcal H^{(1)}$ commutes with the first-order differential operator
\begin{align}\label{KH}
K \mathcal O(z) &=(z^2\partial_z+2z)\mathcal O(z)\,, && [K,\mathcal H^{(1)}]=0\,,
\end{align}
which can be identified with the generator of special conformal transformations along the velocity
direction of the heavy quark, $K=v^\mu K_\mu$.
Since the operators $\mathcal H^{(1)}$ and $K$ commute they can be diagonalized
simultaneously. Thus
eigenfunctions of the (one-loop) evolution kernel coincide with the eigenfunctions
of the differential operator $K$ which are very easy to find (see also~\cite{Bell:2013tfa})
\begin{align}\label{eigenS+}
Q_s(z)=-\frac{1}{z^2} e^{is/z}\,, \qquad s >0
\end{align}
so that
\begin{align}\label{eigenH}
iK\,Q_s(z) =   s \,Q_s(z)\,, && \mathcal{H}^{(1)}\, Q_s(z) = \Big[ \ln(\mu\,s) -\psi(1) -\frac54 \Big]\,Q_s(z)\,.
\end{align}
Alternatively, this result can be written as an operator relation~\cite{Braun:2014owa}
\begin{align}
\label{old}
\mathcal{H}^{(1)}= 4 C_F \ln \left(i\mu e^{2\gamma_E}\, K\right) -5C_F\,,
\end{align}
This (invariant) representation has the same form in momentum and position space.
Once the eigenfunctions and the eigenvalues of the evolution kernel (anomalous dimensions) are
found, the solution of the evolution equation becomes trivial, see~\cite{Braun:2014owa,Bell:2013tfa}
for details.

A new result~\cite{Braun:2019wyx}
which we want to present at this conference is that the representation in~\eqref{old}
can be generalized to all orders in perturbation theory:
\begin{align}\label{kernelHa}
 \mathcal{H}(a) &= \Gamma_{\rm cusp}
 (a) \ln (i \mathcal{K}(a) \mu e^{2\gamma_E}) +
\Gamma_{\!\scriptscriptstyle +}(a)\,,
\end{align}
where $\Gamma_{\rm cusp}(a)$ is the cusp anomalous dimension
 and $\mathcal{K}(a)$ is the generator of special conformal transformations $\mathcal{K}(a) = v^\mu \mathcal{K}_\mu(a)$
in a modified theory: QCD at $d=4-2\epsilon$ at the specially chosen value of the coupling $a_\ast$ such that
$\beta(a_\ast) =0$. $\Gamma_{\!\scriptscriptstyle +}(a)$ is an additive constant that cannot be fixed from symmetry
considerations and requires explicit calculation. This structure was confirmed by explicit calculation to two-loop
accuracy~\cite{Braun:2019wyx}.

In what follows we explain the argumentation that leads to this result and also present explicit expressions for the
eigenfunctions in Mellin space, valid to all orders in perturbation theory, and the corresponding
explicit solution of the evolution equation.

\section{Symmetries of an effective theory}

Before going over to the general case, let us explain how the one-loop result \eqref{old} can be understood on an
intuitive level~\cite{Braun:2018fiz}. It is well known, see e.g. a review in \cite{Braun:2003rp}, that evolution
equations in QCD with massless quarks enjoy conformal invariance at one-loop level: the evolution kernels commute with
the generators of collinear conformal transformations that form an $sl(2)$ algebra. The generators can conveniently be
realized in position space as the first order differential operators acting on quark (gluon) coordinates
\begin{align}\label{generators}
S_+=z^2\partial_{z} + 2 j s,
&&
S_0=z\partial_{z} + {s},
&&
S_-=-\partial_{z}\,,
\end{align}
where $S_-$ corresponds to translations along the light cone, $S_0$ to a certain combination of dilatations and spin
rotation, and $S_+$ to special conformal transformations. Here $j$ is the conformal spin, for quarks $j=1$. Thanks to
the symmetry, the one-loop evolution kernels can be written in the $SL(2)$-invariant form in terms of the corresponding
two-particle quadratic Casimir operator $S_{12}^2 = (\vec S_1 + \vec S_{2})^2$~\cite{Bukhvostov:1985rn}. E.g. for
flavor-nonsinglet leading-twist quark-antiquark operators
\begin{align}
\label{Hik}
  H_{\bar qq} = 4 C_F \biggl\{2 \big[\psi(J_{\bar qq}) -\psi(2)\big] - \frac{1}{J_{\bar qq}(J_{\bar qq}-1)}
   +\frac12\biggr\}\,, && S_{\bar qq}^2 = J_{\bar qq}(J_{\bar qq}-1)\,,
\end{align}
where $\psi(x)$ is the Euler's $\psi$-function.

Physics behind HQET is that one restricts oneself to the situations where the
heavy quark interacts with light particles (quarks and gluons) with momenta that are much smaller than the
quark mass $m_Q$ . In this case the heavy quark becomes almost stationary in its rest frame,
with its wave function oscillating rapidly with time $\Psi(t) \sim e^{-im_Qt}$ so that the generator of translations
reduces to $S_-^{(h)} \sim -im_Q$. This limit can be studied~\cite{Braun:2018fiz} starting from usual $sl(2)$
algebra and rescaling the symmetry generators acting on the heavy quark
\begin{align}\label{rescaling}
     S_-^{(h)}  \to \lambda S_-^{(h)} \,,\qquad  S_+^{(h)} \to \lambda^{-1} S_+^{(h)}\,,
     \qquad \lambda \sim m_Q/\Lambda_{QCD} \to \infty\,.
\end{align}

Consider a  system consisting of one heavy quark $(h)$ and one light (anti)quark $(q)$. The two-particle generators
scale as
\begin{align}
           S_+^{(qh)} &\equiv   S_+^{(q)} +  S_+^{(h)} \mapsto
                S_+^{(q)} +  \lambda^{-1}  S_+^{(h)} =  S_+^{(q)} + \mathcal{O}(\lambda^{-1})\,,
\notag\\
           S_-^{(qh)} &\equiv    S_-^{(q)} + S_-^{(h)} \mapsto
            S_-^{(q)} + \lambda  S_-^{(h)}   =  \lambda S_-^{(h)} + \mathcal{O}(1)\,,
\notag\\
           S_0^{(qh)} &\equiv   S_0^{(q)} +  S_0^{(h)} = \mathcal{O}(1)\,.
\end{align}
The two-particle heavy-light quadratic Casimir operator therefore becomes
\begin{align}
     S_{qh}^2 =    S_+^{(qh)}  S_-^{(qh)} +  S_0^{(qh)}({S_0^{(qh)}}-1) \mapsto
        \lambda  S_+^{(q)}  S_-^{(h)} + \mathcal{O}(1)
\end{align}
and the kernel \eqref{Hik} simplifies to
\begin{align}
  {H}_{qh} =  8C_F  \Big[\psi(J_{qh})-\psi(2)\Big] + \ldots
\mapsto 4 C_F \ln\Big(\lambda S_+^{(q)}  S_-^{(h)} \Big) + \mathcal{O}(\lambda^{-1})\,.
\label{H2-reduction}
\end{align}
Since the ``heavy'' and ``light'' generators act on different spaces we can write, omitting the inessential constant
\begin{align}
    {H}_{qh}  \mapsto  4C_F {\ln\Big(i\mu S_+^{(q)}\Big)} +  4C_F \ln\Big(-i\mu^{-1} \lambda S_-^{(h)}\Big),
\label{facH}
\end{align}
where $\mu$ is an arbitrary parameter with dimension of mass. Thus the heavy and light degrees of freedom decouple
from one another, which is the statement of factorization,  with $\mu$ being the factorization scale.
In HQET only light degrees of freedom remain so that the second term in \eqref{facH} is dropped
and the remaining part
\begin{align}
       {H}_{qh}^{\rm HQET} =  4C_F \ln\Big(i\mu S_+^{(q)}\Big)
\end{align}
coincides with the heavy-light kernel~\eqref{old} (up to the scheme-dependent constant) found in
~\cite{Lange:2003ff,Braun:2003wx,Knodlseder:2011gc} by explicit calculation of one-loop diagrams.
\footnote{The difference between  $K = v^\mu K_\mu \mapsto S_+ = n^\mu K_\mu$ is irrelevant
for the light (anti)quark ``living'' on the light-ray $\sim n^\mu$.}

It is instructive to derive the same result explicitly from the operator algebra. Although QCD interactions preserve
conformal symmetry, the $SL(2)$ invariance of the evolution equations is lost because the heavy-light light-ray
operator \eqref{Oz} does not transform properly under collinear conformal transformations. For this analysis it is
useful to use a representation for the HQET field as a Wilson line in the (negative) $v^\mu$
direction~\cite{Korchemsky:1991zp} so that $\mathcal{O}(z)$ can be viewed as a single (antiquark) operator with
attached Wilson line with light-like and time-like segments and a cusp at the origin. It is easy to check that this
object transforms into itself under infinitesimal (but not finite!) special conformal transformations along $v^\mu$ and
the only source of breaking of dilatation invariance is the appearing of the factorization scale dependent term  $\sim
\ln(i\mu z)$ in the renormalized operator {due to the cusp at the origin}, with the coefficient being (by definition)
the cusp  anomalous dimension $\Gamma_{\rm cusp}(a) = 4C_F a + O(a^2)$. Thus we expect that
\begin{align}
  [K,\mathcal H^{(1)}]=0\,, && [D,\mathcal H^{(1)}]=4C_F
\end{align}
with
\begin{align}
K =(z^2\partial_z+2z)\,, && D=z\partial_z+3/2\,, && [D,K] = K\,,
\end{align}
where the last commutator is part of the usual $sl(2)$ algebra. By virtue of $[K,\mathcal H^{(1)}]=0$ the kernel
$\mathcal H^{(1)}$ must be a certain function of $K$, $\mathcal H^{(1)} = f(K)$. Substituting  $\mathcal H^{(1)}$ by
$f(K)$  in $[D,\mathcal H^{(1)}]=4C_F$ and taking into account  $[D,K]=K$ one obtains a differential equation $K
\partial_K f(K)=4C_F$ so that $f(K) = \ln K +\,\text{\it const}$,  reproducing the result in \eqref{old} (up to the
integration constant).

This argumentation can be generalized beyond the leading order. QCD in non-integer $d=4 -2\epsilon$ dimensions has a
nontrivial critical point \cite{Banks:1981nn} such that $\beta(a_\ast)=0$ for a certain fine-tuned value $a_\ast =
-\epsilon/\beta_0 + \mathcal{O}(\epsilon^2)$ of the coupling. This theory enjoys exact scale and conformal invariance
\cite{Braun:2018mxm} but is very different from ``true'' QCD: In particular it is not asymptotically free as a stable
critical point is only obtained for a large number of flavors. Nevertheless, anomalous  dimensions of composite
operators in this theory in minimal subtraction schemes as functions of the coupling and the number of flavors coincide
exactly with physical QCD at $d=4$ (after one re-expands $\epsilon=\epsilon(a_*)=-\beta_0 a_\ast+O(a_*^2)$) because the
renormalization factors do not depend on $\epsilon$ by construction. Thus the renormalization factors in a physical
theory in the MS schemes have full symmetry of a conformal theory \cite{Braun:2013tva}. In particular we expect that
\begin{align}
\label{exact1}
[\mathcal K(a_\ast),\mathcal H(a_\ast)]=0, && [\mathcal{D}(a_\ast),\mathcal H(a_\ast)] = [D,\mathcal H(a_\ast)]
=\Gamma_\text{cusp}(a_*)\,,
\end{align}
where $\mathcal H(a_\ast)$ is the complete renormalization kernel and $\mathcal D(a_\ast)$, $\mathcal K(a_\ast)$ are
the operators of dilatation and collinear conformal transformations in the interacting theory. They have the following
general form
\begin{align}
\label{exact2}
\mathcal{K} (a_*)&= K  - \epsilon z +  z \Delta(a_\ast) \,, && \mathcal{D}(a_\ast) =  D -\epsilon + \mathcal{H}(a_\ast)\,,
\end{align}
where $\epsilon=\epsilon(a_*)=-\beta_0 a_\ast+O(a_*^2)$. One can show  that the operator $\Delta(a_\ast)$ commutes with
the canonical dilatation operator, $[D,\Delta(a_\ast)]=0$, and therefore
\begin{align}
 [D,\mathcal{K}(a_\ast)] = [\mathcal{D}(a_\ast),\mathcal{K}(a_\ast)]= \mathcal{K}(a_\ast)\,.
\end{align}
 $\Delta(a_\ast)$ is usually referred to as the conformal anomaly.
It can be determined order-by-order in perturbation theory,  $\Delta(a_\ast) = a_\ast \Delta^{(1)} + a_\ast^2
\Delta^{(2)} + \ldots$, from the analysis of conformal Ward identities. A detailed discussion of the relevant
techniques can be found in~\cite{Mueller:1991gd,Belitsky:1998gc,Braun:2016qlg}. The leading (one-loop) contribution
reads~\cite{Braun:2019wyx}
\begin{align}\label{Delta-explicit}
\Delta^{(1)}\mathcal O(z)=C_F \left\{3 \mathcal O(z) +2 \int_0^1 d\alpha\, \omega(\alpha)
\big[ \mathcal O(z)-\mathcal O(\bar\alpha z)\big] \right\},
\end{align}
where
\begin{align}
\omega(\alpha)={2\bar\alpha}/\alpha+\ln\alpha.
\end{align}
Note that $[D,\mathcal H(a_\ast)]= [z\partial_z,\mathcal H(a_\ast)] =\Gamma_\text{cusp}(a_*)$ is  nothing but the
statement that the scale-invariance breaking term, $\ln (i\mu z)$, appears in the evolution kernel $\mathcal H
(a_\ast)$ only linearly with the coefficient $\Gamma_\text{cusp}(a_*)$, \cite{Korchemsky:1987wg}. Using \eqref{exact1},
\eqref{exact2} and following the same line of reasoning as above for the one-loop case one obtains the
representation~\eqref{kernelHa} for the {\it all}-loop kernel.

\section{General solution and two-loop results}

The representation for the evolution kernel $\mathcal H$ in Eq.~\eqref{kernelHa} is very elegant but  for practical
applications one would like to have a more explicit expression as an integral operator,  similar to the one-loop result
in~\eqref{Honeloop}. A general ansatz is
\begin{align}\label{intH}
\mathcal H(a)\mathcal O(z) & =
\boldsymbol{\Gamma}_\text{cusp}(a)\Big[\ln (i\mu_{\scriptscriptstyle\overline{\text{MS}}}\,e^{\gamma_E} z) \mathcal{O}(z)
+
 \int_0^1 d\alpha\frac{\bar\alpha}\alpha\Big(\mathcal{O}(z)-\mathcal O(\bar\alpha z)\Big)
 \notag\\
 &\quad
 + \int_0^1 d\alpha\, \frac{\bar\alpha}\alpha h(a,\alpha)\Big(\mathcal{O}(z)-\mathcal O(\bar\alpha z)\Big)\Big]
 + { \gamma_+(a)}\mathcal{O}(z)\,,
\end{align}
where the expression in the first line is just the one-loop kernel (up to a constant) which commutes with the canonical
generator $K$,   $\gamma_+(a)$ is a constant that requires a separate calculation and the function $h(a,\alpha) = a
h^{1}(\alpha)  + a^2 h^{(2)}(\alpha) +\ldots$ is fixed by the conformal anomaly. Remarkably enough this relation can be
written in a closed form to all orders in the coupling. To this end  we go over to the Mellin representation for the
operator $\mathcal O(z)$,
\begin{align}\label{Mellin}
\mathcal{O}(z)= \int_{-i\infty}^{+i\infty}\! dj\, (i \mu_{\scriptscriptstyle\overline{\text{MS}}}\,
e^{\gamma_E}  z)^j \mathcal O(j)\,.
\end{align}
The evolution kernel \eqref{intH} in this representation becomes
\begin{align}\label{H-mellin}
\mathcal{H}(a) =\boldsymbol\Gamma_\text{cusp}(a)\left[-\frac d{dj}+\psi(j+2)-\psi(2)+\vartheta(j)\right] + \gamma_+(a),
\end{align}
where
\begin{align}
\label{vartheta}
\vartheta(j)={\int^1_0} d\alpha\, \frac{\bar\alpha}\alpha h(\alpha)\Big(1-\bar\alpha^j\Big)\,.
\end{align}
Here and below $\vartheta(j) \equiv \vartheta(a, j)$, $h(\alpha) \equiv  h(a,\alpha)$, etc.

Further, since $[z\partial_z,\Delta(a)]=0$, the conformal anomaly is diagonal on simple powers
\begin{align}
\label{Delta-eigen}
\Delta \, z^j=\delta_j\, z^j\,.
\end{align}
where $\delta_j= a \delta_j^{(1)} + a^2 \delta_j^{(2)}+\ldots$ are the corresponding eigenvalues, and the generator of
special conformal transformations $\mathcal K(a)$ in Mellin representation can be written as a shift operator
\begin{align}
i \mu_{\scriptscriptstyle\overline{\text{MS}}}\,e^{\gamma_E}\,\mathcal K(a) 
    = e^{-\partial_j} \big(j +2 -\epsilon(a)+\delta_j\big)\,,
\end{align}
where $\epsilon(a)= -\beta(a)/a = -\beta_0 a -\beta_1 a^2 -\ldots$. The commutativity condition $[\mathcal K,\mathcal
H]=0$ then translates into a finite-difference equation on  the function $\vartheta(j)$
\begin{align}\label{deltaVj}
\vartheta(j+1) - \vartheta(j) &
=\partial_j\ln \sigma(j), && \sigma(j) = 1+\frac{-\epsilon+\delta_j}{j+2}.
\end{align}
The solution of this equation takes the form
\begin{align}
\vartheta(j)=\vartheta_0 - \partial_j \ln \prod_{k=0}^\infty \sigma(j+k)
=\sum_{k=0}^\infty\left(\frac{\sigma^\prime(k)}{\sigma(k)}-\frac{\sigma^\prime(k+j)}{\sigma(k+j)}\right),
\end{align}
where $\vartheta_0=\sum_{k=0}^\infty{\sigma^\prime(k)}/{\sigma(k)}$ is determined by condition $\vartheta(j=0)=0$.
Observing that $\vartheta(j)$ as defined in \eqref{vartheta} is essentially a Mellin transform of the function
$h(\alpha)/\alpha$ we obtain
\begin{align}\label{h-j-rep}
h(\alpha) & =\int_{-i\infty}^{+i\infty}
\frac{dj}{2\pi i} \,\frac{\vartheta(j+1)-\vartheta(j)}{\bar\alpha^{j+2}}
=\int_{-i\infty}^{+i\infty}
 \frac{dj}{2\pi i} \, \frac{\partial_j\ln \sigma(j)}{\bar\alpha^{j+2}}
=\ln\bar\alpha\int_{-i\infty}^{+i\infty}
 \frac{dj}{2\pi i} \, \,
\frac{\ln \sigma(j)}{\bar\alpha^{j+2}}\,,
\end{align}
where the integration contour goes along the imaginary axis to the right of all singularities of $\vartheta(j)$. This
is the desired result: explicit expression for the evolution kernel in terms of the Mellin integral over the
eigenvalues of the conformal anomaly.

Using the one-loop expression for the anomaly in Eq.~\eqref{Delta-explicit} one obtains from \eqref{Delta-eigen}
\begin{align}
\delta^{(1)}_j=C_F\left(1 +4\left[\psi(j+2)-\psi(2)+\frac12 \frac{\psi(j+2)-\psi(1)}{j+1}\right]\right).
\end{align}
The integral in~\eqref{h-j-rep} can be taken by residues giving rise to the following expression for $h^{(1)}(\alpha)$
\begin{align}
h^{(1)}(\alpha)= \ln\bar\alpha
\left\{ \beta_0-2C_F\left(\frac32 +\ln\frac\alpha{\bar \alpha} +\frac{\ln\alpha}{\bar\alpha}\right)\right\},
\end{align}
which was confirmed in \cite{Braun:2019wyx} by direct calculation.

Since the evolution kernel $\mathcal H$ in Mellin representation \eqref{H-mellin} is a first order differential
operator, its eigenfunctions $$\mathcal H \,\Psi_\lambda=\gamma_\lambda \Psi_\lambda $$  can be found explicitly:
\begin{align}
\Psi_\lambda(j)& =  \Gamma(j+2) e^{-j (\lambda-\gamma_E+1)}  \,\exp\Big[\int^j_0 \!ds\, \vartheta(s)\Big]
= \Gamma(j+2) e^{-j (\lambda-\gamma_E+1-\vartheta_0)}  \prod_{k=0}^\infty \frac{\sigma(k)}{\sigma(j+k)}.
\end{align}
The corresponding eigenvalues are $$\gamma_\lambda = \boldsymbol\Gamma_\text{cusp}(a) \lambda+\gamma_+(a).$$  It is
easy to  check that $\Psi_\lambda(j)$ is also an eigenfunction of the  conformal transformations,  $i\mathcal
K\,\Psi_\lambda = s_\lambda \Psi_\lambda $ with
 \begin{align}
\ln(\mu_{\scriptscriptstyle\overline{\text{MS}}}\,e^{\gamma_E}s_\lambda)= \lambda -\gamma_E+1 -\vartheta_0\,.
\end{align}
Requiring that the two representations  for the evolution kernel in \eqref{kernelHa} and \eqref{intH} coincide on these
eigenfunctions, one obtains the relation between the constant terms $\boldsymbol \Gamma_+$ and $\gamma_+$
\begin{align}
\gamma_+=\boldsymbol\Gamma_+ + \boldsymbol\Gamma_\text{cusp}( 1 - \vartheta_0).
\end{align}
To the NLO accuracy~\cite{Braun:2019wyx}
\begin{align}
  \vartheta(j) &= a  \vartheta^{(1)}(j) = a \biggl\{(\beta_0-3C_F)\Big(\psi^\prime(j+2)-\psi^\prime(2)\Big)
+2C_F\biggl(\frac1{(j+1)^3}
 \notag\\
&\quad
+\psi^\prime(j+2)(\psi(j+2)-\psi(1))+\psi^\prime(j+1)(\psi(j+1)-\psi(1))-\frac{\pi^2}6\biggr)
  \biggr\},
  \notag\\
  \gamma_+(a) &=
   -a C_F +  a^2 C_F
\biggl\{
4 C_F \left[\frac{21}{8} + \frac{\pi^2}{3} - 6\zeta_3\right]
+ C_A \left[\frac{83}{9} -\frac{2\pi^2}{3} - 6\zeta_3\right]
+ \beta_0\left[\frac{35}{18} -\frac{\pi^2}{6}\right]
\biggr\},
\notag\\
  \vartheta_0&= a  \vartheta^{(1)}_0 = a \left(
  \beta_0 \left(1-\frac{\pi^2}6\right) + C_F \left(\frac{\pi^2}6-3
  \right)
  \right),
  \notag\\
  \Psi_\lambda(j)& = \Gamma(j+2)\exp \biggl\{-j (\lambda-\gamma_E+1-\vartheta_0)
    +
a\Big[(\beta_0-3C_F)(S_{j+1}-1)
    + 2C_F S_{j+1}S_j\Big]\biggr\},
\end{align}
where $S_j=\psi(j+1)-\psi(1)$.

Expansion over the eigenfunctions of the evolution kernel in position or momentum space is
in general not very helpful for the solution of the RG equation beyond one-loop level since
the eigenfunctions start to depend on the coupling. In Mellin space, however, the RG equation takes the form
\begin{align}
 \left(\mu \frac{\partial}{\partial\mu}+\beta(a)\frac\partial{\partial a}
 -\boldsymbol\Gamma_\text{cusp}(a)\frac\partial{\partial j} + V(j,a) \right)
\mathcal O(j,a,\mu)=0\,,
\end{align}
where $V(j,a)=j+\gamma_+(a) + \boldsymbol \Gamma_\text{cusp}(a)\Big(\psi(j+2)-\psi(2)+\vartheta(j)\Big)$, and can be
viewed as a RG equation in a theory with two couplings, $a$ and $j$, with the cusp anomalous dimension playing the role
of the beta-function for the latter.%
\footnote{Interestingly enough, similar RGEs appear in the context of TMD factorization, see \cite{Scimemi:2018xaf}.}
The solution of this RGE takes the standard form
\begin{align}
\mathcal O(j(\mu),a(\mu),\mu)=\mathcal O(j(\mu_0),a(\mu_0),\mu_0){\exp\Big\{-\int_{\mu_0}^\mu \frac{ds}{s}\, V(j(s),a(s)) \Big\}}.
\end{align}
Here $j(\mu)$ is the solution of the equation $\mu\partial_\mu j=-\Gamma_\text{cusp}(a)$:
$j(\mu)=j(\mu_0)-\int_{\mu_0}^\mu (ds/s) \, \Gamma_\text{cusp}(a(s))$ with the boundary condition $j(\mu_0)=j$.

To summarize, we have shown that evolution equations for heavy-light operators (hence LCDAs of heavy-light hadrons)
have an elegant group-theory interpretation such that the evolution kernels can be written as a logarithm
of the generator of collinear conformal transformations in QCD at $d-2\epsilon$ dimensions at the critical point,
up to an additive constant. This relation is exact to all orders in perturbation theory and is verified in
\cite{Braun:2019wyx} by explicit calculation to two-loop accuracy.
The resulting two-loop evolution equation is directly relevant for phenomenology and allows
one, e.g., to perform a complete next-to-next-to-leading logarithmic
(NNLL) resummation of heavy quark mass logarithms in the $B\to \ell \nu_\ell\gamma$ decay.
The corresponding analysis and a discussion of further applications go beyond the subject of this talk.


\vspace*{0.2cm}

{\large\bf Acknowledgments:}~~\\
This work was supported by the DFG, grants BR 2021/7-2 (YJ), MO 1801/1-3 (AM).

\end{document}